\documentclass{aa}
\usepackage{txfonts}
\usepackage{graphicx}
\usepackage{subfigure}
\usepackage{verbatim}

\begin{document}

\title{Spectroscopy of the symbiotic binary CH~Cygni from 1996 to 2007}

\author{M. Burmeister\inst{1,2}
	\and L. Leedj\"{a}rv\inst{1}}
                                                                                
\offprints{M. Burmeister, \email{ mari@aai.ee}}
                                                                                
\institute{Tartu Observatory, 61602 T\~{o}ravere, Estonia
	\and Institute of Physics, University of Tartu,
	T\"{a}he 4, 51010 Tartu, Estonia}
                                                                                
\date{Received 20 January 2009 / Accepted 2 June 2009}

\abstract
{Efforts to uncover the nature of the variable star CH~Cyg have lasted for
decades by now. Recently, the model of a red giant and an
accreting white dwarf on a $\sim$5700-day orbit has seemed to have gained an
advantage over other ideas, mainly as a result of the investigation of
the cool component's absorption lines. In addition to those lines the
star's spectrum also presents emission features, which 
form near the hot component and are therefore a probe of the
accretion phenomenon, which in turn could be bound to the orbital
motion.}
{We have monitored a set of emission lines in
the optical spectra of CH~Cyg over an extended period of time
to determine relations between the behaviour of those lines and the
photometric data. We are searching for a possible connection with the star's
orbital motion, and find hints of
the physical phenomena occurring near the white dwarf.}
{More than 500 observations of CH~Cyg were carried out during $1996-2007$
using the 1.5-m telescope at the Tartu Observatory, Estonia.
Most of the spectra were registered in the H$\alpha$ region,
but other Balmer lines and lines of He, N, O, and Fe were also
investigated in terms of their equivalent widths, radial velocities,
and absolute fluxes.}
{The spectra indicate different stages that CH~Cyg has been through
in the course of our observations. During quiescence, the strength
of the line profiles changes in opposite phase with the star's luminosity.
The H$\alpha$ profile becomes double-peaked at times,
which may point to the temporary presence of some disk-like structure,
but also to absorption
in neutral gas surrounding the area of the formation of the emission
component. Occasionally
the line develops wide wings, which indicate high velocities of
the matter.
In 1999, when a periastron passage is assumed according to the
5700-day model,
the Balmer lines strengthen considerably, as might be expected due to 
an increase in the accretion rate.
The eclipse in 2001, however, is not noticed.
At certain episodes, the controversial 756-day period of CH~Cyg is
seen in our data.} 
{}


\titlerunning{Spectroscopy of CH~Cygni}
\maketitle

\section{Introduction}
\label{sect:introduction}

CH~Cygni is a variable star that has been studied extensively for
a long time and over a wide spectral range (see Miko\l ajewski et al.\
\cite{miko90a} for a historical review).
The star has for some time been included in the group of
symbiotic stars and is one of the brightest among them with
its average visual magnitude of about $7.1$ (Belczy\'{n}ski et al.\
\cite{belczynski}). It is also one of the nearest symbiotics with a
distance of $268\pm62$\,pc measured by \emph{Hipparcos}
(Munari et al.\ \cite{munari97}).

For a long time, however, CH~Cyg was considered
a single red giant experiencing low-amplitude 
pulsations with a period around
100 days. It did not show any spectral peculiarities
and was even adopted as a standard star of the spectral
class M6~III in the Morgan-Keenan classification,
and yet in 1963 a strong blue continuum and
\ion{H}{I} emission lines appeared in the star's spectrum
(Deutsch \cite{deutsch64}).
At that time, the spectrum of CH~Cyg resembled a symbiotic star.
This phenomenon was observed again in 1965.
Since then, there have been several such outbursts of different durations:
$1967-1970$, $1977-1986$, $1992-1995$, and
$1998-1999$, separated by quiescent periods, during which the spectrum of
CH~Cyg resembled that of
a single red giant, like before 1963.
According to its infrared colours, CH~Cyg is classified as
an S-type symbiotic. However, its
orbital period appears to be unusually long for that class, and
sometimes it shows strong
[\ion{N}{II}] emission lines that are more characteristic of the D-type
systems.

During the longest and most pronounced outburst in $1977-1986$, a sudden
drop in visual light was recorded in 1984, followed by the formation of a
jet detected at radio wavelengths (Taylor et al.\ \cite{taylor86}). In 1996,
after a minor outburst, the deepest minimum of visual light
($V\sim10^{\mathrm{m}}$) ever recorded took place (e.g., Miko\l ajewski et
al.\ \cite{miko96}). A radio map obtained by Karovska et
al.\ (\cite{karovska98}) in January 1997 revealed an elongated structure,
which in their opinion could be related to jet formation. Crocker et al.\
(\cite{crocker}) found jet-like extensions in their radio images obtained in
1999 during the $1998-1999$ outburst.

CH~Cyg is one of the few symbiotics demonstrating flickering on a
timescale of minutes (Dobrzycka et al.\ \cite{dobrzycka}).
Simultaneously with the appearance of the jets,
the disappearance of
the flickering and the enhancement of the blue continuum has been observed,
as well as sudden broadening of emission lines.

About ten symbiotic systems have demonstrated jets by now
(Brocksopp et al.\ \cite{brocksopp}; Leedj\"{a}rv \cite{leed04}).
The mechanism of jet formation is not clear, but it is
suspected that it is related to
an accretion disk around the compact component. In this case,
some symbiotics may have at least transient accretion disks.
The disappearance of flickering after jet formation could then
point to the disruption of the disk (Sokoloski \& Kenyon \cite{soko03}).

It is generally believed that the activity of CH~Cyg is
powered by the energy generated by the accretion of the
red giant wind by the hot component
as the luminosity of the latter
(a few hundred $L_\odot$ at the highest) seems to be too
low to be caused by a nuclear-burning shell.
The amount of the accreted matter may depend
on the pulsations of the giant, as well as on the orbital
motion of the stars, if the orbit is elliptical.

When it comes to precise details of the system, no
generally accepted view exists, although several possible models
have been presented.
A magnetic rotator model was proposed by Miko\l ajewski et al.\
(\cite{miko90b}).
According to them, CH~Cyg contains a pulsating red giant and
a white dwarf with a strong magnetic field
on an eccentric long-period orbit.
Miko\l ajewski et al.\ (\cite{miko90a}) propose an orbital
period of 5700 days.

Hinkle et al.\ (\cite{hinkle93})
discovered very regular variations with a 756-day period from the
radial velocities in the infrared spectra of CH~Cyg, which
they interpreted as the orbital period of the symbiotic
binary. Their model also involved a G dwarf as a third component,
orbiting the binary with a 5300-day period.
This model found both improvement
(Skopal \cite{skopal95}; Skopal et al.\
\cite{skopal96b}) and opposition
(Munari et al.\ \cite{munari96}). The last authors suggested that
the 756-day
modulation is caused by the pulsation of the giant, rather than by
orbital motion. In light of new observations and the improved
estimates of system parameters, Hinkle et al.\ (\cite{hinkle09}) renounce
their triple-star model and suggest an orbital period of 5689 days for the
symbiotic pair, close to the periods in former binary models.

We present our spectroscopic observations of
CH~Cyg at Tartu Observatory from 1996 to 2007.
In Sect.~\ref{sect:observations} we describe our observational equipment,
the data,
reduction tools, and measurements.
The overview of the star's photometric and spectroscopic variations
follows in Sects~\ref{sect:photometry} and
\ref{sect:variations}, respectively.
Analysis of the results is presented in Sect.~\ref{sect:analysis}.


\section{Observations and data reduction}
\label{sect:observations}

All the spectra treated in this paper
were obtained at Tartu Observatory using the
1.5\,m telescope equipped with the Cassegrain grating spectrograph.
Until 1999 March, the detector was
the SpectraSource Instruments
CCD camera HPC-1 (Tek $1024\times1024$ chip, pixel size
$24\times24\,\mu$m, Peltier cooled).
From 1999 March to 2006
March, the spectra were recorded with the cryogenically
(liquid nitrogen) cooled CCD camera Orbis-1 of the same
company (Tek $512\times512$ chip, pixel size
$24\times24\,\mu$m). Since then we have used the
Andor Technologies CCD camera Newton DU-970N ($400\times1600$
chip, pixel size $16\times16\,\mu$m, Peltier cooled).

From 1996 to 2007,
almost 500 spectra were obtained, which cover the time interval
without major gaps.
In general, the spectra can be divided into two categories: red and blue.
Most of the red spectra were made  with
an 1800 lines-per-mm (lpm) grating and
linear dispersion of either 0.29\,\AA\,px$^{-1}$ (HPC-1 and Orbis-1 cameras),
or 0.19\,\AA\,px$^{-1}$ (Newton camera).
An 1200 lpm grating was used for the rest of the spectra in the
red region, which therefore have lower resolution.
All the spectra in the blue region were made with the 1200 lpm grating.
Table~\ref{table:regions} lists the wavelength coverages with
corresponding lines for each camera.
\begin{table}
\caption{Wavelength regions, covered by our spectra with different
cameras.}
\label{table:regions}
\centering
\begin{tabular}{lll}                                                 \hline
  \bf{HPC-1}          &  \bf{Orbis-1}        & \bf{Newton DU-970N} \\
                      &                      &                     \\
  H$\alpha$\,(1800)   &  H$\alpha$\,(1800)   & H$\alpha$\,(1800)   \\
  $6500-6750$\,\AA    &  $6490-6620$\,\AA    & $6475-6740$\,\AA    \\
                      &                      &                     \\
  H$\alpha$\,(1200)   &  H$\alpha$\,(1200)   & H$\alpha$\,(1200)   \\
  $6200-6900$\,\AA    &  $6230-6620$\,\AA    & $6130-6900$\,\AA    \\
                      &                      &                     \\
  H$\beta$            &  H$\beta$            & H$\gamma$-H$\beta$  \\
  $4200-5050$\,\AA    &  $4610-5040$\,\AA    & $4200-5120$\,\AA    \\
                      &                      &                     \\
                      &  H$\delta$-H$\gamma$ &                     \\
                      &  $3930-4400$\,\AA    &                     \\ \hline
\end{tabular}
\end{table}
The high-resolution red spectra contain
the H$\alpha$,
[\ion{N}{II}] 6548, and 6584 lines. Velocities of these lines could
be measured with an accuracy of $3-4\,\mathrm{km\,s^{-1}}$
(HPC-1 and Orbis-1)
or $2.5-3\,\mathrm{km\,s^{-1}}$ (Newton). The accuracy was lower
for the rest of the lines, 
$\sim$15$\,\mathrm{km\,s^{-1}}$ at worst. The
[\ion{O}{I}] 6300 and 6364 lines are found in the red spectra with
lower resolution.
The blue spectra made with Orbis-1 camera include either H$\beta$ with
[\ion{O}{III}] 4959 and 5007 lines or H$\gamma$,
[\ion{O}{III}] 4363, and H$\delta$ lines. Blue spectra by
Newton DU-970N contain all these lines, except H$\delta$.
Certain lines of neutral helium ($\lambda\lambda$ 4026, 4388,
4471, and 6678) fall into some of our spectra as well.

The spectra were reduced using the software package MIDAS provided
by ESO, except some earlier spectra, for which a software
package KASPEK, developed at Tartu Observatory, was applied.
After subtracting the dark frame (in the case of
HPC-1 spectra) or bias (Orbis-1 spectra) and sky background,
the spectra were calibrated to the wavelength scale using an
Ne-Ar (until 1999 October) or Th-Ar (since 1999 October)
hollow cathode lamp. The calibrated spectra were normalised
to the continuum.
The following properties of emission lines were measured:
peak intensity, equivalent width, velocity, and
full width at half maximum, of which the first three are
treated in this paper.

\emph{Remark.}
In the figures of this paper we also present some data that have
been obtained before the period under discussion, $1996-2007$.
Their purpose is illustrative; no high-resolution spectra
suitable for more precise investigation can be found among them.


\section{Photometric variability of CH~Cygni}
\label{sect:photometry}

\begin{figure}
\resizebox{\hsize}{!}{\includegraphics[angle=0]{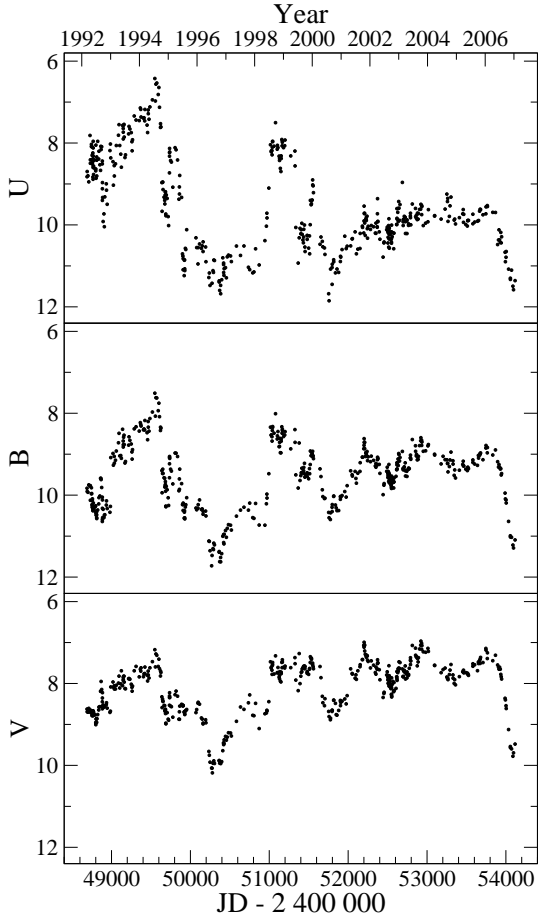}}
\caption {Recent $U$, $B$, and $V$ data of CH~Cyg,
all on the same scale for easier comparison.}
\label{fig:ubv}
\end{figure}

As we did not perform any photometry ourselves, the
photometric data necessary for calculating the absolute fluxes of lines
(Sect.~\ref{subsect:fluxes}) were gathered from the literature
(Skopal et al.\
\cite{skopal95jt},
\cite{skopal96a},
\cite{skopal00},
\cite{skopal02},
\cite{skopal04},
\cite{skopal07};
Skopal
\cite{skopal97},
\cite{skopal98};
Hric et al.\ \cite{hric96};
Leedj\"{a}rv \& Miko\l ajewski \cite{leed00}).
The $UBV$ values are presented in Fig.~\ref{fig:ubv}.
We do not possess any observations after January 2007, but the visual
estimates by AAVSO show that the $V$ brightness of the star
that faded to 10th magnitude by the end of 2006 and remained on that
low level throughout 2007.
According to Taranova \& Shenavrin (\cite{taranova06}), the fading
was particularly remarkable in the infrared spectral region, as by
the end of 2006 the $J$ magnitude of CH~Cyg was about 0.7 magnitudes
fainter than during the previous, similar event that occurred in 1996.
Possibly a new rise started in October 2007.

Photometric observations have shown the luminosity of CH~Cyg to be
variable on both short and long timescales.
Several periods have been found from the stars' light curves,
though none of them is obvious.
The shortest of these periods (minutes or hours) are generally associated
with the activity of the hot component and the longer ones
(starting from 100 days) with the pulsation and rotation of
the giant, as well as the orbital motion of the binary.
Among the longer periods, those of 756 and 5700 days have gained
special interest, especially as
similar periods have been found
from the red giant absorption features (Hinkle et al.\ \cite{hinkle93}).
The shortest one of these resembles the
orbital periods of other symbiotic stars, but is not
always present in the light curve.
The period of $\sim$5700 days, first reported by Yamashita \&
Maehara (\cite{yamashita79}), is possibly related to
the orbital motion, although this is exceptionally long
for an S-type symbiotic star.


\section{Variations in the spectrum: equivalent widths and radial
velocities}
\label{sect:variations}

\begin{figure}
\resizebox{\hsize}{!}{\includegraphics[angle=0]{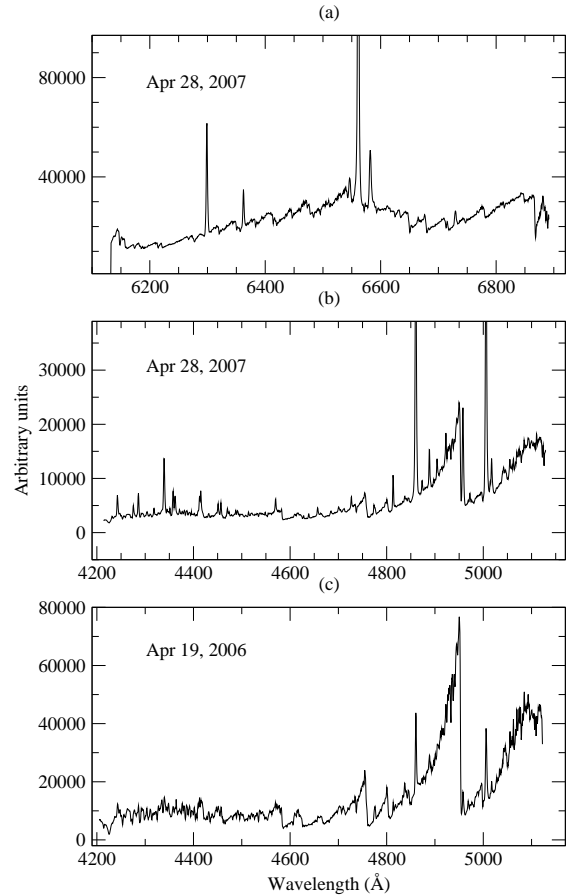}}
\caption {Representative lower-resolution spectra from
2007 April 28, when the continuum was weak and the emission
lines exceptionally conspicuous (\emph{a} and \emph{b}),
and from 2006 April 19 (\emph{c}), when the emission lines were weak.
Most of the time of our observations the spectra of CH~Cyg looked
like the latter.}
\label{fig:HaHbReg}
\end{figure}

Symbiotic stars are described as having emission lines in their
spectra, but the prominence of these lines varies from system to
system (see the spectra of different symbiotics in Munari \& Zwitter
\cite{munari02}). The emission line spectrum of CH~Cyg is relatively
undistinguished.
Most of the time, only the Balmer lines are
easily found; the lines of neutral helium are very weak and
high excitation lines tend to be missing. In the red part of the
optical spectrum, molecular absorption of the M giant can be seen.
However,
the emission spectrum becomes more conspicuous
at certain episodes.
As our observations cover the interval $1996-2007$, they include
the last outburst in $1998-1999$ and the more or less quiescent time
since then. Balmer lines were strong during the
outburst and the light minima in 1996 and $2006-2007$, with strong forbidden
lines seen only during the mentioned minima.
Figure~\ref{fig:HaHbReg} shows representative spectra from
red and blue regions, made with Newton DU-970N.

\subsection{Balmer lines}
\label{subsect:balmer}

\begin{figure*}
\centering
\includegraphics[angle=270,width=17cm]{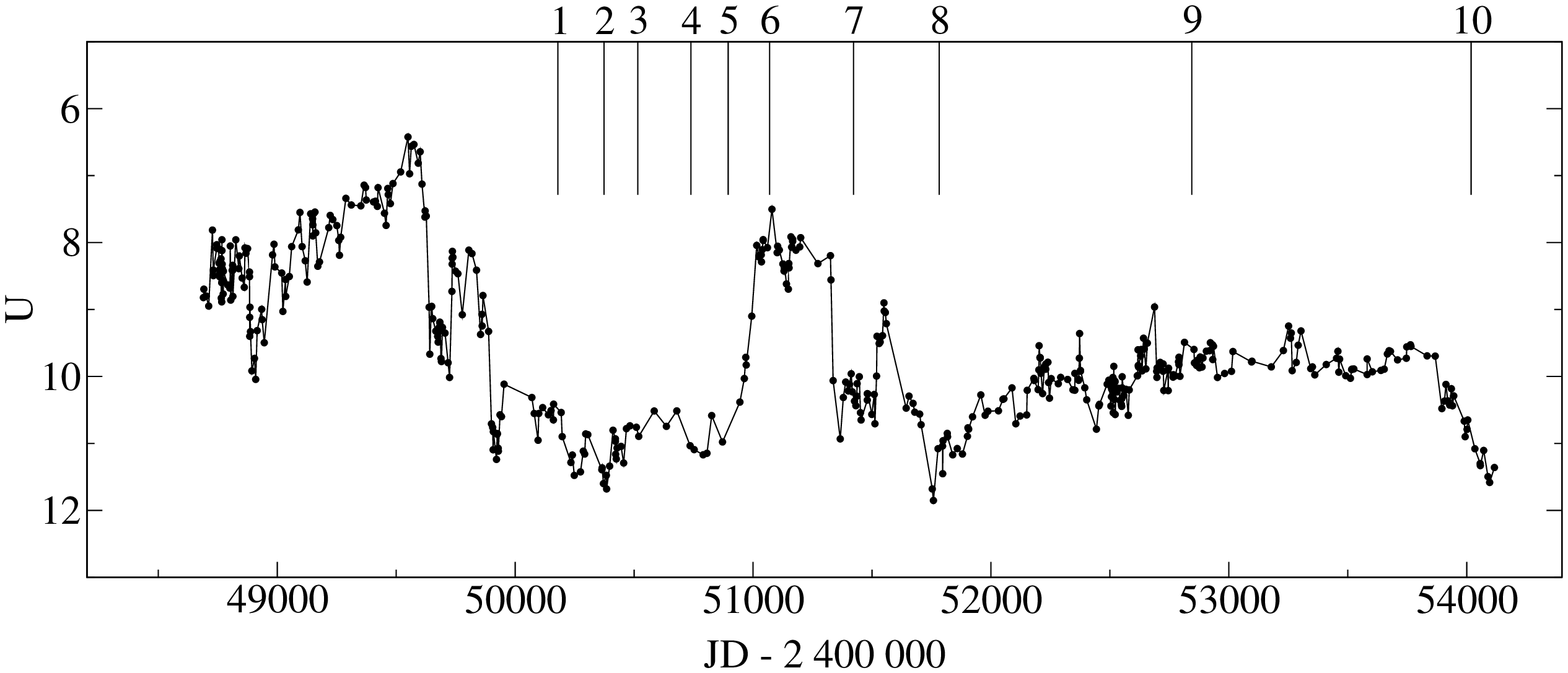}
\includegraphics[angle=270,width=17cm]{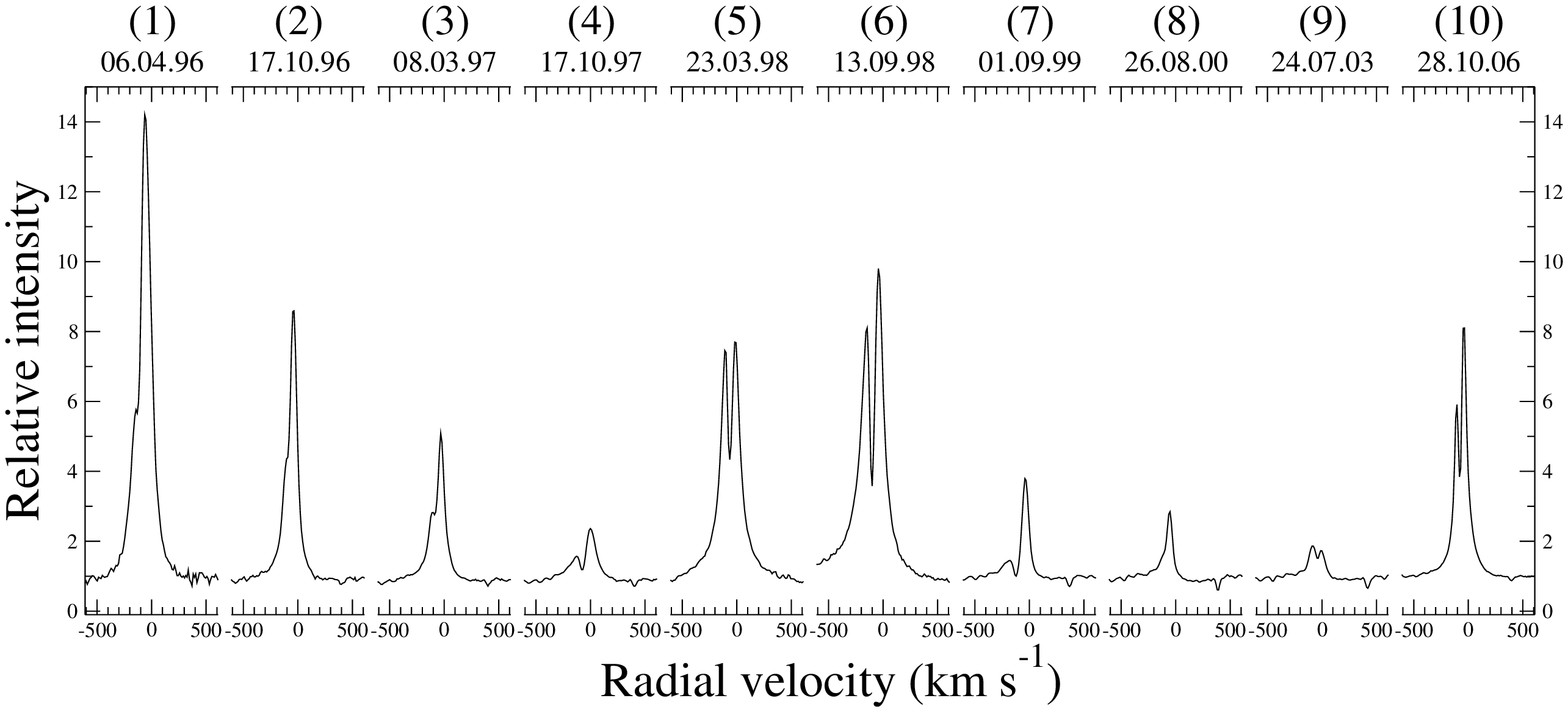}
\caption{\emph{Top:} Recent $U$ light curve of CH~Cyg. Numbered positions
correspond to the spectra below. \emph{Bottom:}
Some examples of the H$\alpha$ profile of CH~Cyg, all plotted
on the same scale for comparison.}
\label{fig:profiles}
\end{figure*}

Our main interest was concentrated on the H$\alpha$ line, the strongest
of the emission lines in CH~Cyg spectrum. Most of the spectra in
the region of that line were made with a resolution high enough to allow
the investigation of the fine structure of the line.
During the time interval under consideration,
the star was mostly quiescent ($U \geq 10^{\mathrm{m}}$), however, activity occurred
before 1996 and in $1998 - 2000$.
The intensity and shape
of the H$\alpha$ emission line were highly variable, and sometimes the
line became double-peaked. Figure~\ref{fig:profiles} shows some
examples of the H$\alpha$ line profile at different times.
When the line is double-peaked,
the two emission components often have very different
heights, the red peak generally being stronger than the blue one.
Single-peaked line profiles are
usually asymmetric with a step on the blue side,
alluding to a blue-shifted absorption component.
These shapes can occur in cases of both weak and strong lines.

Below we characterise the H$\alpha$ line variations
between 1996 and 2007, using equivalent width (EW) as a measure of its
strength, and relying mostly on the spectra taken with the
1800 lpm grating (see Sect.~\ref{sect:observations}).
After the qualitative description of the line profiles' evolution, the
quantitative evaluation of the radial velocities is given.

The spectra in early 1996 showed
a strong H$\alpha$ line, which had a single-peaked profile
with a slight step on the blue side
(see the first line in Fig.~\ref{fig:profiles}).
During the second half of the year,
the line was considerably weaker (second line), but had a similar shape.

In 1997 EW of the line continued to vary, but was on average lower than
the year before. The shape of the profile remained almost
unchanged, that is,
single-peaked and asymmetric. For some time in
October when the line was very weak (fourth
line in Fig.~\ref{fig:profiles}),
the step on the blue side deepened into an absorption component.

In early 1998 the H$\alpha$ line was still weak, but
developed wide wings (fifth line in Fig.~\ref{fig:profiles})
by March, i.e.\ before the increase in brightness, which started
in the middle of the year.
In August and September the
line was at its highest intensity (sixth line in Fig.~\ref{fig:profiles})
and at the same time asymmetric, with
the blue wing extending to
velocities of $-1100$\,km\,s$^{-1}$. The red wing,
in contrast, remained relatively moderate with a
velocity of about 400\,km\,s$^{-1}$.
The profile had two peaks of
comparable strength for most of the year,
but the blue peak declined
during the last months.

\begin{figure}
\resizebox{\hsize}{!}{\includegraphics[angle=270]{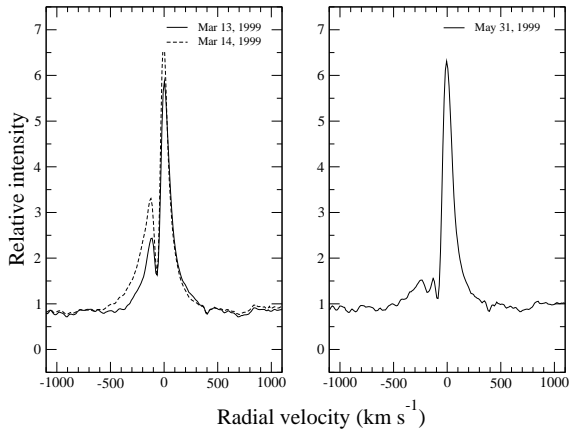}}
\caption{\emph{Left:} Spectra from two consecutive nights to illustrate the
rapid changes of the H$\alpha$ line in spring 1999.
\emph{Right:} Profile with two absorption components.}
\label{fig:kevad99}
\end{figure}

Rapid changes took place in 1999.
During the first months, the H$\alpha$ line was still
strong and double-peaked, varying rapidly at the same time
(Fig.~\ref{fig:kevad99}, left panel).
Multiple absorption components in the blue wing, seen already in 
December 1998, became more prominent by May 1999.
On May 31, two absorption components were clearly visible,
as shown in Fig.~\ref{fig:kevad99} (right panel). Their velocities
proved to be $-95$\,km\,s$^{-1}$ and
$-180$\,km\,s$^{-1}$.
In summer 1999, the red peak also became considerably weaker.
Starting from November 1999, the H$\alpha$ line was
very weak
for a long time.
In December 1999, the absorption component was very deep.
At the beginning of 2000, the line had two relatively equal
peaks, after which the blue peak weakened and
disappeared by summer. The line remained single-peaked,
though asymmetric with a variable shape (eighth
line in Fig.~\ref{fig:profiles}), until the
end of 2002. Then it became stronger and
developed an absorption component in the centre.
During 2003 and 2004 the line was weak and double-peaked
(ninth line in Fig.~\ref{fig:profiles}),
alternatively double-peaked and single-peaked
during 2005.
In 2005 it was again somewhat stronger
and from August 2006 kept strengthening
until the beginning of 2007. The shape of the line
was double-peaked with the red peak more intense than the blue
one (tenth line in Fig.~\ref{fig:profiles}).
At that time, the visual
luminosity of the star reached its minimum at about 10 magnitudes.
In spring 2007, the H$\alpha$ line started to decline and
reached the $2004-2006$ state by the end of the year.

The resolution of our H$\beta$ spectra was too low to
investigate the shape of the line in any detail. The double-peaked profile
in 1998, for instance, observed by Eyres et al.\ \cite{eyres02},
was not resolved. The measured EWs, however,
behaved similarly to those of H$\alpha$.
Once in a while, we also made spectra in the region of
H$\gamma$ and H$\delta$. These lines are present mainly during
active times. No H$\delta$ emission can be found in our spectra since
May 1999, and H$\gamma$ was at times recogniseable as a very weak
red-shifted emission. These appearances tend to be in phase
with the maxima of the H$\alpha$ and H$\beta$ line strengths.

\begin{figure}
\resizebox{\hsize}{!}{\includegraphics{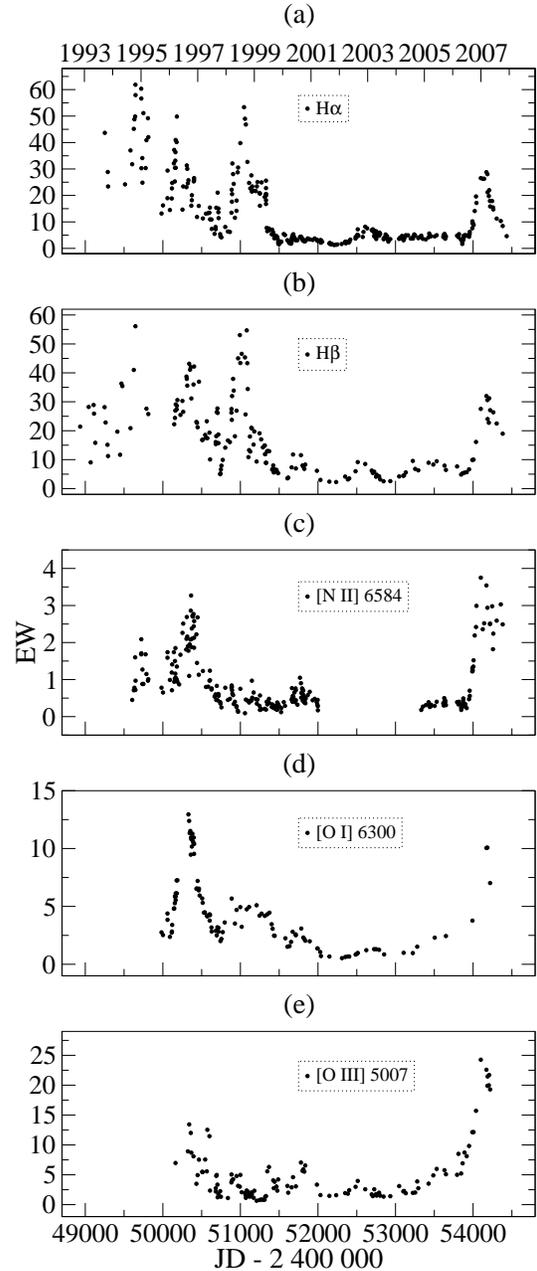}}
\caption{EWs of some emission lines of CH~Cyg.}
\label{fig:5EW}
\end{figure}

The measured EWs of H$\alpha$ and H$\beta$ are depicted in
Figs~\ref{fig:5EW} (a) and (b), respectively.
It can be seen that, during quiescence, the EWs of H$\alpha$ and
H$\beta$ lines
change in a wave-like manner. This is even more visible in
peak intensities (see Sect.~\ref{subsect:short_per}).

\begin{figure}
\resizebox{\hsize}{!}{\includegraphics[angle=270]{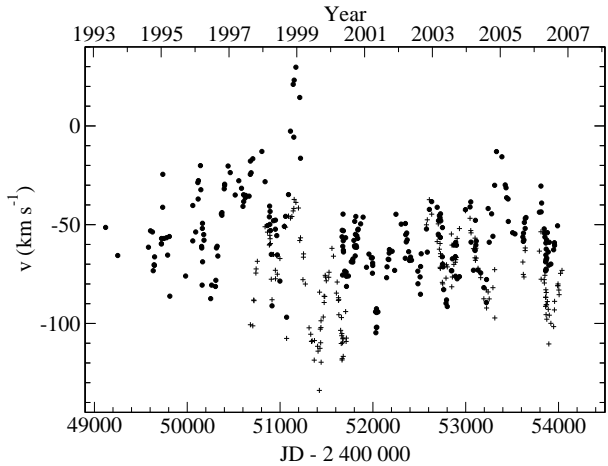}}
\caption{Velocities of the H$\alpha$ emission line (filled circles)
and its absorption component (crosses).}
\label{fig:hakiirused}
\end{figure}

\begin{figure}
\resizebox{\hsize}{!}{\includegraphics[angle=270]{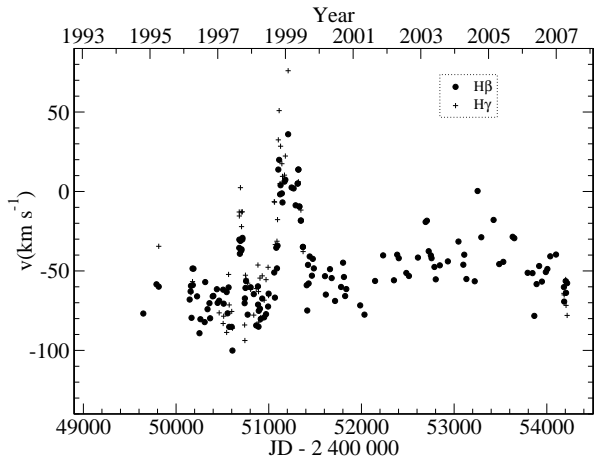}}
\caption{ Velocities of H$\beta$ and H$\gamma$.}
\label{fig:vhbhg}
\end{figure}

We tried to measure the radial velocities of the H$\alpha$
line of CH~Cyg by fitting a Gaussian profile to it, though
the procedure was often (especially in 1999)
complicated by the asymmetric shape
of the line. 
One Gaussian was fitted to the emission line and another to the
absorption component (if present). As seen from Fig.~\ref{fig:hakiirused},
both velocities vary with a period of one year. The cause of such
periodicity remains a puzzle to us.
Velocities of the H$\beta$ and H$\gamma$ are plotted in
Fig.~\ref{fig:vhbhg}.
Generally they accumulate around the system's radial velocity
($-59.9\,\mathrm{km\,s^{-1}}$ according to Hinkle et al.\
\cite{hinkle09}),
except during a time interval from autumn 1998 to spring 1999,
when they
appear to be significantly more positive
than usual. This is caused by the very
asymmetric profiles of the
Balmer lines at that time. The blue-shifted absorption components
devoured the blue part of the line and let the red part dominate.
The one-year wave does not seem to be present here.

According to the ephemeris by Hinkle et al.\ (\cite{hinkle09}),
the CH~Cyg system passed the periastron on JD\,2\,451\,370
(July 1999). Although the proposed eccentricity ($e=0.122$)
is rather small, it could be sufficient to enhance mass transfer
from the red giant to the white dwarf, therefore 
strengthening the Balmer emission lines. This effect is seen
especially well in the absolute fluxes (Fig.~\ref{fig:5Flux}).
However, the H$\alpha$ and H$\beta$ lines, as well as the [\ion{O}{I}]
6300, line reached the maximum intensity a little before the proposed
periastron passage, at a phase around 0.95.

\subsection{Helium lines}
\label{subsect:helium}

In contrast to most of the symbiotic stars, helium lines are usually
weak or absent in the spectrum of CH~Cyg.
As found by Leedj\"{a}rv \& Miko\l ajewski (\cite{leed00}), the lines
of neutral helium $\lambda\lambda$4471, 4388, and 4026 are only visible
when the star is brighter than about
$10^{\mathrm{m}}$ in the $U$ band. This also applies to our later spectra.
In late 1998, when the hydrogen Balmer lines were strong and double-peaked,
the above-mentioned \ion{He}{I} lines developed P~Cygni profiles.
In some of our red spectra, \ion{He}{I} 6678 was visible
as a faint single-peaked emission.
For a short period in $1998-1999$ its EW increased
to $1\,\AA$ or more.

\begin{figure}
\resizebox{\hsize}{!}{\includegraphics[angle=270]{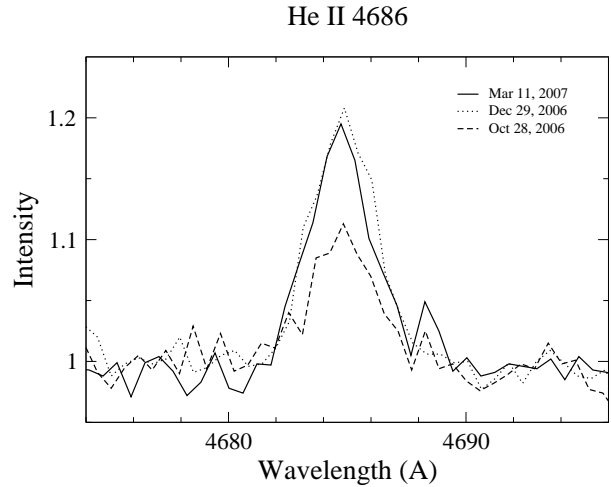}}
\caption{\ion{He}{II} 4686 line.}
\label{fig:4686}
\end{figure}

There have only been a few incidents when the
\ion{He}{II} 4686 line, so characteristic of many symbiotics stars, was seen
in the spectrum of CH~Cyg. In 1984 it was discovered as a very wide
emission feature by Leedj\"{a}rv et al.\ (\cite{leed94}).
In the present spectra, the line was visible
for about 100 days in 1998, at the time of brightness
maximum ($U\sim8^{\mathrm{m}}$).
It appeared for the second time in October 2006,
and had at least doubled its strength by the end of the year.
In 2007 March 11, when our next spectrum in the blue region was made.
The profile of the helium line was much the same (Fig.~\ref{fig:4686}).
The line became weaker by summer and was no longer to be found in
the spectrum made in October 2007. Interestingly, this
appearance of \ion{He}{II} 4686 took place during a brightness
\emph{minimum}. $UBV$ photometry shows a steep decrease in light in all bands
since the summer of 2006; by the beginning of 2007, the
$U$ magnitude had reached $\sim$11.5$^{\mathrm{m}}$. We possess no $UBV$ data
since that, but the visual AAVSO light curve indicates a new rise from the
minimum in the middle of 2007, just about when the helium line disappeared.

\subsection{[\ion{N}{II}] lines}
\label{subsect:nitrogen}

As a rule, the lines of [\ion{N}{II}] 6548 and 6584 tend to be absent in
S-type symbiotics, but present in D-type systems
(Van Winckel et al.\ \cite{winckel93}).
In our spectra of CH~Cyg, these lines were visible
most of the time.

The behaviour of the EW of [\ion{N}{II}] 6584 (the more distinguished of the
two) is shown in Fig.~\ref{fig:5EW} (c). The line was present and quite
strong in 1996, but decreased in 1997. It disappeared completely and quite
suddenly in spring 2001 and was missing until the end of 2004, when it
became barely visible again. During the following year the line was very
weak, but detectable. It remained at about the same level until August 2006,
when it started to increase steeply. At the end of 2006 and at the beginning
of 2007 it was about as strong as in 1996.

The behaviour of [\ion{N}{II}] 6548 was similar to 6584.
In 1998 the [\ion{N}{II}] 6548 line became overwhelmed by
the extended blue wing of H$\alpha$.

The profiles of the [\ion{N}{II}] lines are complicated, especially when
the lines are weak, so their velocities are more meaningful when measured
at times when the line is stronger. In $1996-1997$
the velocities were more negative than in $2006-2007$.
In the later
period the velocities were consistent with those of Balmer lines,
about $-60\,\mathrm{km\,s^{-1}}$, which is close to the systemic velocity.
During $1996-1997$, however, the velocities were about
$-100\,\mathrm{km\,s^{-1}}$.
H$\alpha$ was stronger in $1996-1997$
than it was in $2006-2007$, but not so wide as to reach the
nitrogen lines, so we do not think it had any effect on their
velocities.

\subsection{[\ion{O}{I}], [\ion{O}{III}], and other forbidden lines}
\label{subsect:oxygen}

The forbidden lines
[\ion{O}{I}] 6300 and 6364 can be found in our low-resolution red spectra.
The EWs of these lines seem to follow roughly the course of the Balmer lines
(those of [\ion{O}{I}] 6300 are depicted in Fig.~\ref{fig:5EW} (d)),
except that they have a major maximum during 1996 and the outburst in
1998 is unremarkable.

Our spectra in the H$\beta$ region include [\ion{O}{III}] 4959 and
[\ion{O}{III}] 5007. The earliest and latest spectra made with
HPC-1 and Andor cameras also cover the region of [\ion{O}{III}] 4363,
and so do the H$\gamma$--H$\delta$ region spectra.
This line, however, is not to be found in any of our spectra.
[\ion{O}{III}] 5007, however, is always present,
although
it was quite weak
from 2001 to 2005.
Measured EWs of [\ion{O}{III}] 5007 are given in Fig.~\ref{fig:5EW} (e).
[\ion{O}{III}] 4959 becomes clearly visible in 2005, remaining
much weaker than 5007.
As this line is located just at the edge of the TiO 4955 absorption
band, it is difficult to normalise this spectral region to the
local continuum level. For that reason, the measured quantities may be less
trustworthy than those of 5007 and we do not present them here, but
they do seem to follow the course of those of 5007 line.

On some of our spectra, [\ion{S}{II}] 4069 line can be found,
as well as [\ion{S}{II}] 6731.
Its strength appears to behave similarly to [\ion{O}{III}] lines.
Like [\ion{N}{II}] lines, [\ion{S}{II}] lines are also usually not found
in the spectra of S-type symbiotic stars.

\subsection{Ionised Fe lines}
\label{subsect:ferrum}

Our blue spectra made by the Orbis camera span about 4610\,\AA\ to
5040\,\AA. This spectral region includes many permitted and forbidden
lines of singly ionised iron, like \ion{Fe}{II} 4629, [\ion{Fe}{II}] 4640,
[\ion{Fe}{II}] 4729, [\ion{Fe}{II}] 4815, [\ion{Fe}{II}] 4874, [\ion{Fe}{II}]
4890, [\ion{Fe}{II}] 4905, \ion{Fe}{II} 4924, and \ion{Fe}{II}
5018, of which the last two are blended with \ion{He}{I}.

In May 1999, most of the listed lines were clearly visible, with
the permitted lines stronger than the forbidden ones.
Beginning from June, the \ion{Fe}{II} lines were mainly gone,
except 5018, but as said, it is not clear how much of a contribution
the line gets from \ion{He}{I}. [\ion{Fe}{II}] lines
were not much changed at first and did not begin to fade until autumn.

During the following years, both \ion{Fe}{II} and [\ion{Fe}{II}]
lines were very weak or missing. From the autumn of 2002 we
only have one blue spectrum made in November, and on that
several Fe lines seem to have strengthened. The intensity of
the \ion{Fe}{II} lines appeared to reach a local maximum
at the beginning of 2003 and that of the [\ion{Fe}{II}] lines
somewhat later, in April. The \ion{Fe}{II} lines did not
become as strong as they were in May 1999, but some of the
[\ion{Fe}{II}] lines
even exceeded the intensities they had in 1999.

During 2004, the iron features were rather weak, although the \ion{Fe}{II}
lines strengthened for a short time in September. During the following year,
\ion{Fe}{II} lines were basically changeless, [\ion{Fe}{II}] seemed
to show some maximum in September.

Significant changes took place in the year 2006. The visual brightness
of CH~Cyg started to decline in summer, and the intensity of iron
lines grew considerably from September on. By the end of the year,
the intensities of [\ion{Fe}{II}] lines were 3--4 times of their
intensities in May 1999. The \ion{Fe}{II} lines did not grow
to such a degree as to exceed the intensities in May 1999.

In 2007, both \ion{Fe}{II} and [\ion{Fe}{II}] became step by step weaker.

\subsection{Absolute fluxes}
\label{subsect:fluxes}

The appearance of emission lines depends on the level of the continuum:
when the continuum is modest, the lines rise more into view; with a
strong continuum, they seem weakened.
For that reason, conversion of the relative intensity into the
absolute fluxes is necessary
for obtaining the true picture of line strengths.
We gathered the necessary photometric data in $B$, $V$, and $R$ bands
from the literature (see Sect. 3).

\begin{figure}
\resizebox{\hsize}{!}{\includegraphics{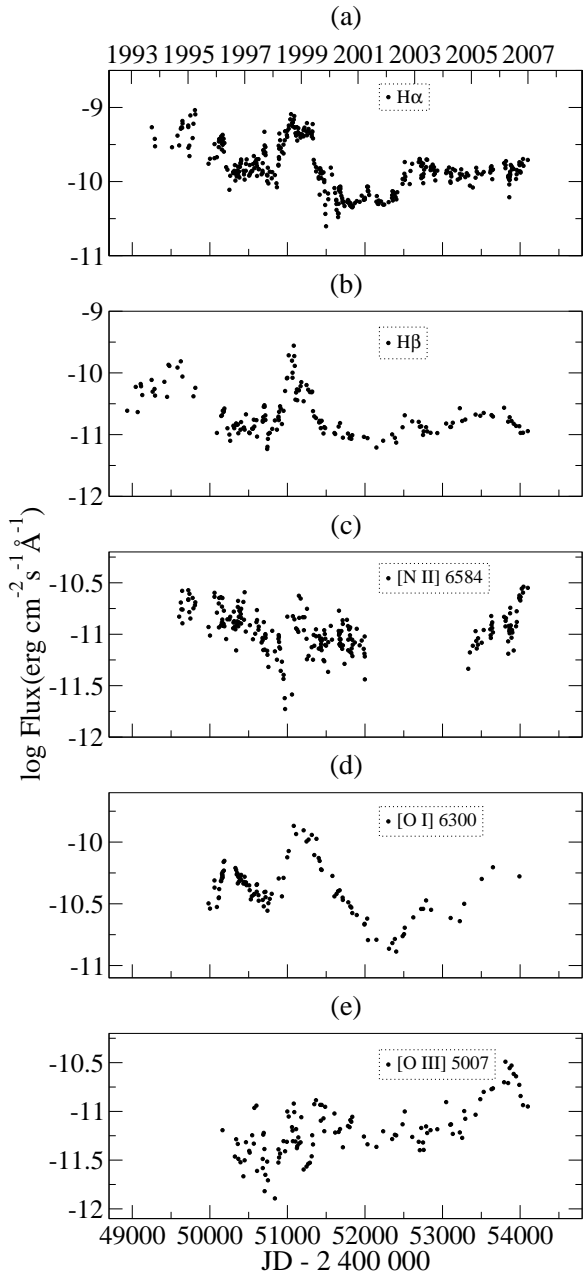}}
\caption{Fluxes of some emission lines of CH~Cyg.}
\label{fig:5Flux}
\end{figure}

As the dates of photometric data usually did
not coincide with the dates of our spectra, the values of
$B$, $V$, and $R$ magnitudes were
interpolated. When no $R$ values were available, these
were computed from $V$ values after finding an approximate relation
between $R$ and $V$ magnitudes: $R = 0.744\times V - 1.287$.
Photometric data were corrected for interstellar extinction using
the colour excess $E(B-V)=0.07$ taken
from Slovak \& Africano (\cite{slovak78}). The computed fluxes are given
in Fig.~\ref{fig:5Flux} for the same lines as in Fig.~\ref{fig:5EW}.

The outbursts that CH~Cyg had in $1992-1995$ and $1998-1999$
are reflected in the fluxes and EWs of H$\alpha$, as well as in the
star's light curves.
In 1996, when the luminosity has a minimum
($V\sim10^{\mathrm{m}}$), the EW, in
contrast, has a maximum (recall the strong single-peaked profiles
in 1996).
The same happens in 2006, when the luminosity suddenly decreases,
EW increases, and the flux, as a result, remains unchanged.
This tendency can also be seen in the corresponding figures of
H$\beta$. In the case of [\ion{N}{II}] 6584 line, the outbursts
do not echo in the EWs, but the minimum-maximum correspondence
of continuum and lines in 1996 and 2006 exists here as well.
As a consequence, the flux is nearly constant all the time,
although the dispersion of the points is large.
Our [\ion{O}{I}] 6300 data begin at the light maximum of 1996,
during which EW of the line becomes high. During the $1998-1999$
the EWs only have a small maximum, but the flux increases
noticeably.


\section{Analysis and discussion}
\label{sect:analysis}

\subsection{Origin of the H$\alpha$ line}
\label{subsect:originha}

The interpretation of the H$\alpha$ line profile depends on how we
assume the line is formed.
Balmer lines of symbiotic stars are generally thought to originate
in that part of the red giant wind that is ionised by the
hot component.
The double-peaked structure that is often seen in the Balmer
lines of symbiotics (van Winckel et al.\ \cite{winckel93};
Ivison et al.\ \cite{ivison94})
is probably caused by self-absorption in the
neutral hydrogen region around the ionised area. The
absorption component tends to be shifted blueward from the
line centre.
This is caused by the increasing velocity of the wind --
the absorbing particles in the neutral part of the
wind are moving faster towards us
than the emitting particles in the underlying ionised region.
The depth of the absorption component indicates
the density of the wind.

Asymmetric
H$\alpha$ profiles with a blueshifted absorption
of different depths
prevail during our observations.
The line had a similar shape during the $1967-1970$ outburst
(Faraggiana \& Hack \cite{faraggiana}) and $1992-1995$
outburst (Skopal et al.\ \cite{skopal96a}).
However, during the longest outburst in $1977-1986$
the ratio of the peaks of the H$\alpha$ line was mostly
reversed, as reported by Hack et al.\ (\cite{hack86}).
Robinson et al.\ (\cite{robinson}) propose that this
effect could be caused by the elliptical orbit of the star.

In some symbiotic systems
(particularly T~CrB and CI~Cyg, according to
Robinson et al. \cite{robinson}), the double-peaked line profiles
could also arise from the accretion
disk around the white dwarf, like in the case of cataclysmic variables.
It probably cannot be excluded
that at times a transient accretion disk may form around the
hot component of CH~Cyg and give rise to the sometimes observed
double-peaked profiles.

\subsection{Possible accretion disk in 1998}
\label{subsect:disk}

\begin{figure}
\resizebox{\hsize}{!}{\includegraphics[angle=270]{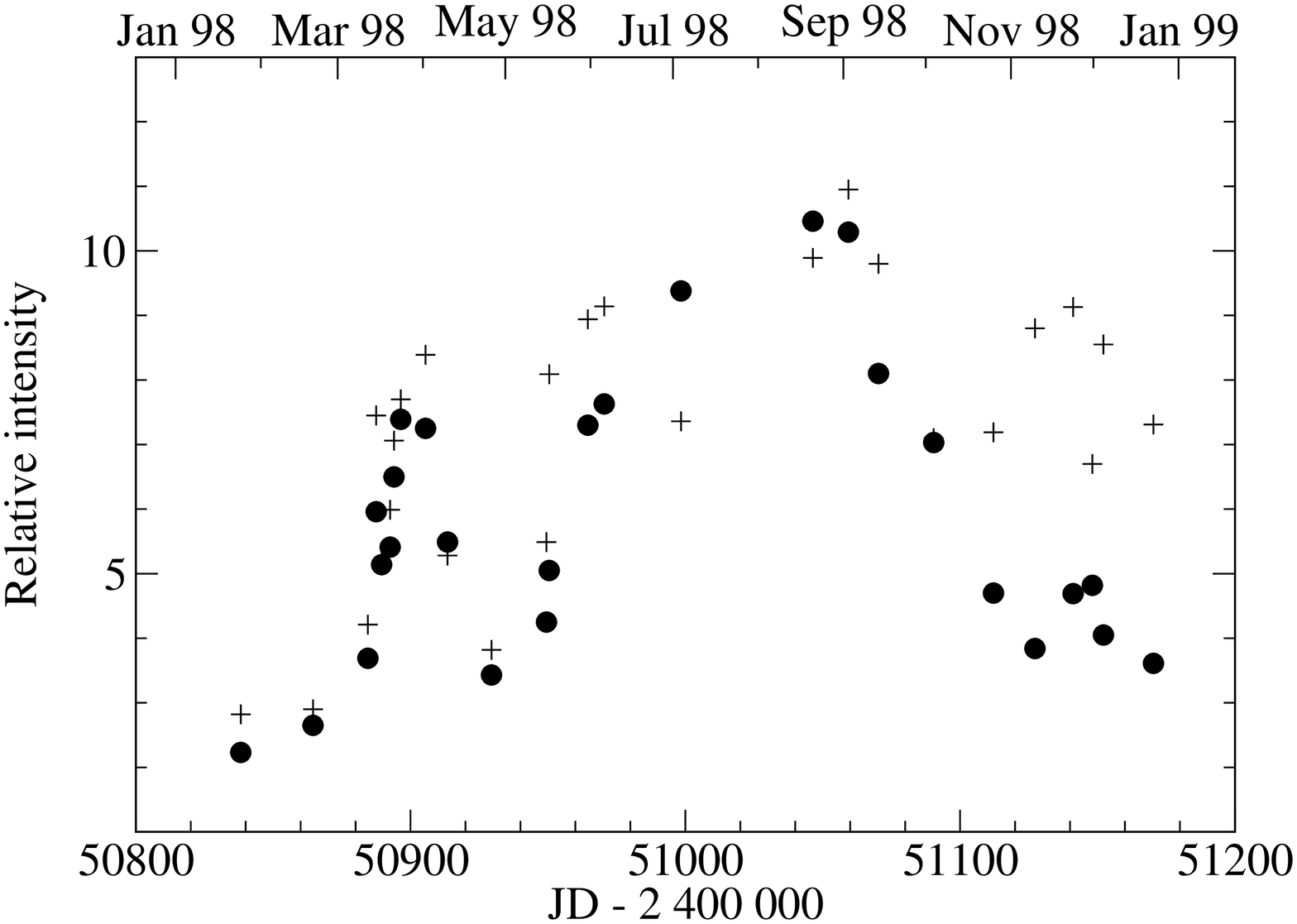}}
\resizebox{\hsize}{!}{\includegraphics[angle=270]{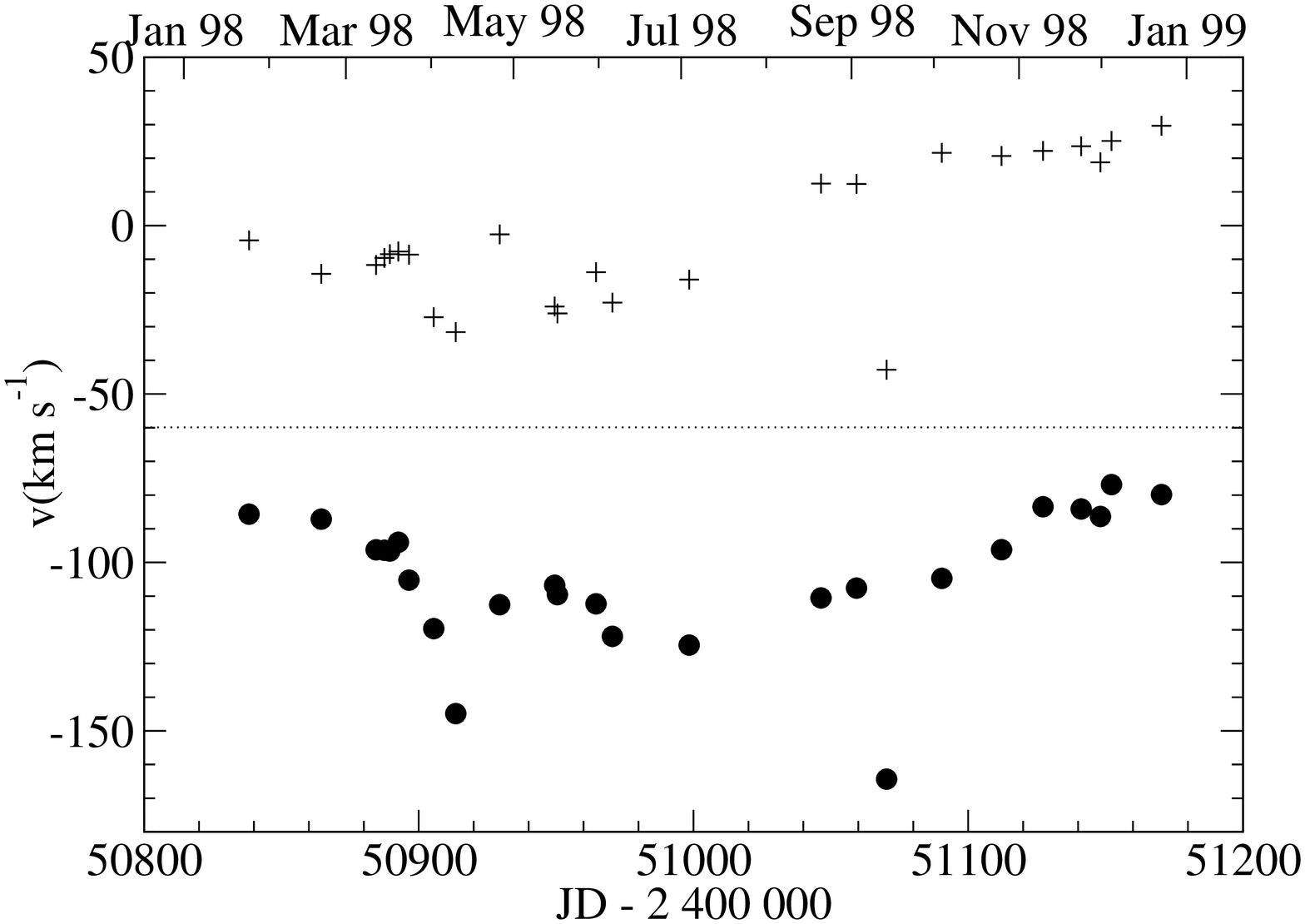}}
\caption{\emph{Top:} Relative intensities (with respect to the continuum) of
the blue and red peaks of H$\alpha$ in 1998.
\emph{Bottom:} Radial velocities of the two peaks. On both graphs, filled
circles denote the blue peak and crosses the red peak. The dotted line
denotes the systemic velocity.}
\label{fig:2Tops}
\end{figure}

We conjecture that, during our observations, the accretion disk
could have existed in 1998, when
the H$\alpha$ profile was double-peaked
with almost equal peaks (Fig.~\ref{fig:2Tops}, upper panel) and
had notably wide wings.
Fitting the peaks with two Gaussians showed that
their positions with regard to the systemic radial velocity
varied somewhat, as can be seen from the lower panel of
Fig.~\ref{fig:2Tops}.
The depth of the central absorption was about half the height of
the peaks at that time.
The idea of the line's origin in a disk is supported by the high
time resolution photometry
by Sokoloski \& Kenyon (\cite{soko03}), who found
exceptionally high-amplitude flickering in the light curves
of CH~Cyg in 1998 and also proposed an accretion disk as the source
of these. At the same time, the H$\alpha$ profiles were similar
to those in 1982, for which the disk interpretation was successfully
used (Leedj\"{a}rv et al.\ \cite{leed94}).

The method of Doppler tomography has been useful for interpreting
of the emission lines from accretion
disks of cataclysmic variables, but the very long
orbital period of CH~Cyg and a short-lived presence of its accretion
disk make little use of it here.
Nonetheless, we tried to make some estimations of the
parameters of the possible disk, based on some
assumptions used in Doppler tomography (e.g., Marsh \& Horne
\cite{marsh88}; Robinson et al.\ \cite{robinson}).
The peak-to-peak separation in the line profile
in velocity units is assumed to be
equal to twice the radial velocity of the material orbiting at
the outer edge of the disk.
If we know the inclination of the disk and the mass of
the accreting star, then
from that velocity it is possible
to derive the outer radius of the disk 
(see Leedj\"{a}rv et al.\ \cite{leed94} for details).

\begin{figure}
\resizebox{\hsize}{!}{\includegraphics[angle=270]{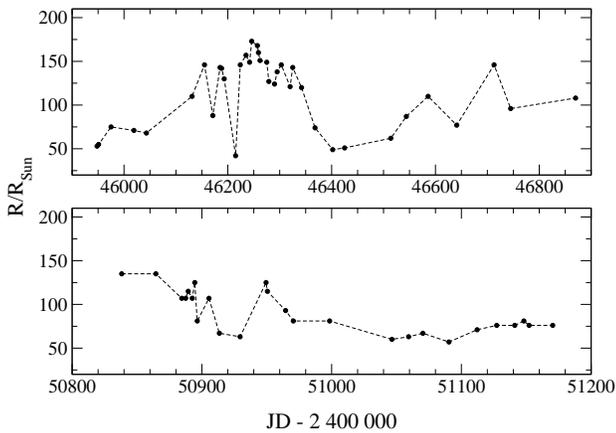}}
\caption{Radii of the hypothetic accretion disk around the
hot component of CH~Cyg in $1984-1987$ (upper panel) and
in 1998 (lower panel).}
\label{fig:radius}
\end{figure}

We assumed the accretion disk to lie in the orbital plane of the
system. The system inclination of CH~Cyg is not known, but
has to be high enough to allow eclipses.
Following the example of Robinson et al.\ (\cite{robinson}),
we adopted the value $i=78^{\circ}$. The mass of the accreting
star $M=1\,M_{\odot}$ was taken from Miko\l ajewski
\& Miko\l ajewska (\cite{miko88}).
As a result we found that
the value of the outer radius falls in the range
$(4-9) \times 10^{12}$\,cm, or
$50-150\,R_{\odot}$.
The radial velocities of the peaks in $1984-1987$ Leedj\"{a}rv et al.\
(\cite{leed94}) were used to estimate the outer radius by the same method
and are brought together with our results in Fig.~\ref{fig:radius}.
The values of the outer radius seem to flock around $70-75\,R_{\odot}$
in both graphs, with the exception of the interval
JD\,2\,446\,150-350. The results obtained at that time
might be an artefact due to the eclipse of the hot component that
must have taken place then according to the 5700-d model
and to the ephemeris
\[JD(periastron)=2\,445\,681(\pm 192)+5689.2(\pm 47.0)\times E\]
by Hinkle et al.\ (\cite{hinkle09}).

\subsection{Variations with the 2.1 year period}
\label{subsect:short_per}

A 2.1-yr period of unclear origin has been repeatedly noticed in the
behaviour of CH~Cyg.
Miko\l ajewski et al.\ (\cite{miko90a}) report a $\sim$770-day
periodicity after analysing the long-term optical and near-infrared
light curves; $3-5$ deep minima separated by $700-800$ days are also found
from $1986-1991$ multicolour photometry by Miko\l ajewski et al.\
(\cite{mikolajewski92}). Hinkle et al.\ (\cite{hinkle93}) present a 756-day
period as the orbital one of the symbiotic system in their triple-star
model. Hinkle et al.\ (\cite{hinkle09}) confirm the existence of a 2.1-yr
period and discuss its two most probable sources: a light companion of the
red giant or its nonradial pulsation.

\begin{figure}
\resizebox{\hsize}{!}{\includegraphics[angle=270]{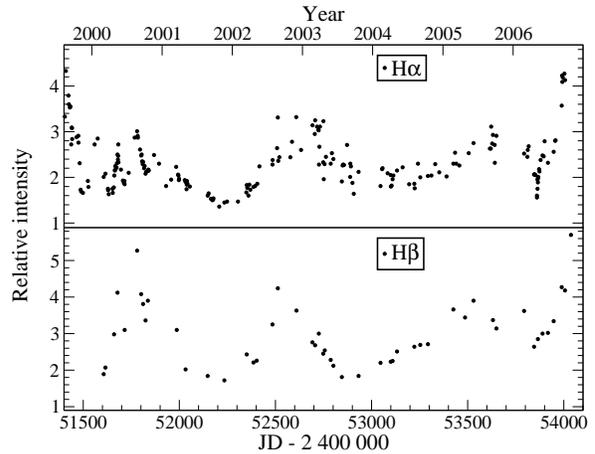}}
\caption{Intensities of the H$\alpha$ and H$\beta$ lines
with respect to the continuum} during quiescence.
\label{fig:HaHbI}
\end{figure}

Signs of this period can be found in our spectra.
Figure~\ref{fig:HaHbI} shows the peak intensities of
H$\alpha$ and H$\beta$ in the time span between the summer
of 1999 and the end of 2006, when the lines were weak.
Either line shows wavelike behaviour, whereby the minima of the waves
correspond to the maxima of the star's light curves, so that the H$\alpha$
and H$\beta$ lines are the strongest when the luminosity of the star is the
lowest. Visual inspection suggests the period of this variation to be about
2 years. We searched for periodicities from that part of the peak intensity
curves and also the corresponding part of EW curves of H$\alpha$ and
H$\beta$ using phase dispersion minimisation methods. The applied program
was ISDA, developed by Dr J.~Pelt in Tartu Observatory (see
http://www.aai.ee/$\sim$pelt/soft.htm\#ISDA). Of the several methods
applied, only that of Stellingwerf gave consistent results, which indeed
fell near 756 days. The scatter of the points was large, however.

\subsection{H$\alpha$ lines of CH~Cyg and other symbiotics}
\label{subsect:othersymbs}

Based on the high-resolution spectra of 59 southern symbiotic and related stars,
van Winckel et al.\ (1993) divided both S- and D-type symbiotics into three
subtypes according to the shape and width of their H$\alpha$ profiles.

Stars of type S-1 have single-peaked asymmetric H$\alpha$ lines, and those of
type S-2 possess a shallow, and those of type S-3 deep absorption components.
The H$\alpha$ lines of D-1 type stars are narrow, single-peaked, and quite
symmetric, the lines of type D-2 are similar but have a broader base, and type
D-3 H$\alpha$ profiles are broad and have an absorption component in the
middle.

It is hard to decide into which subtype CH~Cyg fits. Its H$\alpha$ profile
seems to change between S-1 and S-2 subtypes, but the lack or weakness of
other lines that characterises CH~Cyg is rather typical of type S-3. On the
other hand, we have seen that CH~Cyg sometimes presents rather strong
[\ion{N}{II}] lines in its spectra. This fact could suggest its belonging to
D-type, because these lines are almost never found in the spectra of S-type
symbiotics and are very common among D-type systems.

Substantial changes in the shape of the H$\alpha$ profile have been detected
in a few other symbiotics.
Van Winckel et al.\ find that in both S- and D-types of stars, the width of
the H$\alpha$ line and the depth of its absorption component correlate
with the strength of the He lines and the visibility of the nebular lines.
In the case of CH~Cyg, the strength of other lines changes in phase with the
H$\alpha$ line, but the existence of an absorption component does not seem
to play much of a role.

\section{Conclusions}
\label{sect:conclusions}

As a consequence of observing the EWs of the Balmer emission
lines of CH~Cyg between
1996 to 2007, three maxima were noted: in 1996, in $1998-1999$,
and in $2006-2007$. The first of these was also seen in
the EWs of the forbidden lines of [\ion{N}{II}], [\ion{O}{I}],
and, to a lesser extent, [\ion{O}{III}]. Comparison with simultaneous
photometry revealed, however, that the star's light had a minimum
at that time. Therefore, the strengthening of the
lines was deceptive, because it was only due to the weakness of the
underlying continuum. The second maximum, on the other hand,
coincided exactly with the rise and fall of the $U$ light.
It only appeared in the EWs (and fluxes) of Balmer lines;
no sign of any strengthening was seen in the forbidden lines.
The $2006-2007$ maximum in its turn was most conspicuous in the EWs of
the forbidden lines. Like the 1996 event, it also took place
during brightness minimum. The fluxes of Balmer lines show no
maximum, those of the forbidden lines, however, are remarkably high.
The [\ion{O}{III}] lines have been previously noticed as
accompanying mass outflows, on account of what we suspect the presence
of extended emissions at that time.
Regrettably
there have so far been no reports of radio mapping since 2000, infrared
imaging since 2003, or optical imaging since 1999.

Intensive accretion to the white dwarf was presumably responsible for
the 1998 event. The outburst was observable in the Balmer lines, which are
supposed to form near the accretor, but undetected in the forbidden
lines, arising in a tenuous environment farther away.
The strong double-peaked profile of the H$\alpha$ line, supported
by observations of flickering, suggests the presence of
an accretion disk around the white dwarf at that time.
The outer radius of such a disk was estimated to be about $70-75\,R_{\odot}$.
The disk was likely disrupted in 1999, when the H$\alpha$ and
H$\beta$ lines rapidly decayed and flickering disappeared
(Eyres et al.\ \cite{eyres02}).
The similarity of the H$\alpha$ profiles in $1982-1983$ (Leedj\"{a}rv
et al.\ \cite{leed94}) and $1998-1999$ seems to confirm the orbital
period $5689^{\mathrm{d}}$ and an elliptical orbit proposed by
Hinkle et al.\ (\cite{hinkle09}). Both episodes of symmetric
double-peaked profiles took place slightly before the periastron
passage.

For more than half the period of our observations
(in particular between 2000 and 2006), both the
luminosity and the strength of the emission lines remained on an
approximately constant, low level.
The star did not completely lose its symbiotic features, though,
as the [\ion{O}{I}] and [\ion{O}{III}] lines remained visible.
The 756-day period, which has previously been seen in the infrared velocities,
was found in our Balmer lines' intensities at that time.
It is possible that this wave, like the increases in
1996 and $2006-2007$ is caused by the changes in the underlying
continuum; in this case, however, it is unclear why its amplitude
is even slightly larger in H$\beta$ data.

There are a few more symbiotic stars known whose hot components
have a luminosity below or around $100\,L_{\odot}$, e.g. MWC~560,
R~Aqr, EG~And, BX~Mon. The first two are well-known jet sources,
while both CH~Cyg and R~Aqr also have shown X-ray jets (Kellogg
et al.\ \cite{kellogg07}). Shugarov et al.\ (\cite{shugarov07})
link V407~Cyg to these objects and propose that
a new subclass of symbiotic stars
could exist with CH~Cyg, V407~Cyg,
and R~Aqr as prototypes. A characteristic feature of those stars
would be an accretion disk around the white dwarf, which gives
rise to flickering on short time scales and occasionally causes the ejection
of jets. Most likely those stars
have longer orbital periods than do S-type symbiotic stars.


\begin{acknowledgements}
The authors wish to thank K.~Annuk, T.~Eenm\"{a}e, A.~Hirv, and
A.~Puss, who did a large part of the observations.
The work was partly supported by the Estonian Science Foundation
grant No. 6810, and by the target-financed project SF0060030s08
financed by the Ministry of Education and Research of Estonia.
\end{acknowledgements}


%

\begin{thebibliography}{99}
\bibitem[2000]{belczynski} Belczy\'{n}ski,~K., Miko\l ajewska,~J.,
  Munari,~U., Ivison,~R.~J., Friedjung,~M. 2000, A\&AS, 146, 407
\bibitem[2006]{biller06} Biller,~B.~A., Close,~L.~M., Li,~A. et al.
  2006, ApJ, 647, 464
\bibitem[2004]{brocksopp} Brocksopp,~C., Sokoloski, J.~L.,
  Kaiser,~C. et al. 2004, MNRAS, 347, 430
\bibitem[2001]{crocker} Crocker,~M.~M., Davis,~R.~J., Eyres,~S.~P.~S.,
  et al. 2001, MNRAS, 326, 781
\bibitem[1964]{deutsch64} Deutsch,~A.~J. 1964, Ann. Rept. Mt. Wilson
  and Palomar Obs. 1963, 11
\bibitem[1996]{dobrzycka} Dobrzycka,~D., Kenyon,~S.~J., Milone,~A.~E.
  1996, AJ, 111, 414
\bibitem[2002]{eyres02} Eyres,~S.~P.~S., Bode,~M.~F., Skopal,~A.
  et al. 2002, MNRAS, 335, 526
\bibitem[1998]{ezuka98} Ezuka,~H., Ishida,~M., Makino,~F. 1998, ApJ,
  499, 388
\bibitem[1971]{faraggiana} Faraggiana,~R., Hack,~M. 1971, A\&A, 15, 55
\bibitem[2004]{galloway04} Galloway,~D.~K., Sokoloski,~J.~L. 2004,
  ApJ, 613, L61
\bibitem[1986]{hack86} Hack,~M., Rusconi,~L., Sedmak,~G. et al.
  1986, A\&A, 159, 117
\bibitem[1993]{hinkle93} Hinkle,~K.~H., Fekel,~F.~C., Johnson,~D.~S.,
  Scharlach,~W.~W.~G. 1993, AJ, 105, 1074
\bibitem[2009]{hinkle09} Hinkle,~K.~H., Fekel,~F.~C., Joyce,~R.
  2009, ApJ, 692, 1360
\bibitem[1996]{hric96} Hric,~L., Skopal,~A., Urban,~Z., et al. 1996,
  CoSka, 26, 46
\bibitem[1994]{ivison94} Ivison,~R.~J., Bode,~M.~F., Meaburn,~J.
  1994, A\&AS, 103, 201
\bibitem[1998]{karovska98} Karovska,~M., Carilli,~C.~L., Mattei,~J.~A.
  1998, JAVSO, 26, 97
\bibitem[2007]{karovska07} Karovska,~M., Carilli,~C.,~L., Raymond,~J.~C.,
  Mattei,~J.~A. 2007, ApJ, 661, 1048
\bibitem[2007]{kellogg07} Kellogg,~E., Anderson,~C., Korreck,~K. et al.
 2007, ApJ, 664, 1079
\bibitem[1994]{leed94} Leedj\"{a}rv,~L., Miko\l ajewski,~M.,
  Tomov,~T. 1994, A\&A, 287, 543
\bibitem[2000]{leed00} Leedj\"{a}rv,~L. Miko\l ajewski,~M. 2000,
 ASPC, 204, 345
\bibitem[2004]{leed04} Leedj\"{a}rv,~L. 2004, BaltA, 13, 109
\bibitem[1988]{marsh88} Marsh,~T.~R., Horne,~K. 1988, MNRAS, 235, 269
\bibitem[1988]{miko88} Miko\l ajewski,~M., Miko\l ajewska,~J.
  1988, in The Symbiotic Phenomenon,
  ed. J.~Mikolajewska, M.~Friedjung, S.~J.~Kenyon, R.~Viotti, IAU Coll.
  103. Reidel, Dordrecht, p. 233
\bibitem[1990a]{miko90a} Miko\l ajewski,~M., Miko\l ajewka,~J.,
  Khudyakova,~T.~N. 1990a, A\&A, 235, 219
\bibitem[1990b]{miko90b} Miko\l ajewski,~M., Miko\l ajewska,~J., Tomov,~T.,
  Kulesza,~B., Szczerba,~R. 1990b, AcA, 40, 129
\bibitem[1992]{mikolajewski92} Miko\l ajewski,~M., Miko\l ajewska,~J.,
  Khudyakova,~T.~N. 1992, A\&A, 254, 127
\bibitem[1986]{mikolajewski86} Miko\l ajewski,~M., Tomov,~Y. 1986,
  MNRAS, 219, 13p
\bibitem[1996]{miko96} Miko\l ajewski,~M., Tomov,~T.~V., Kolev,~D.,
  Leedj\"{arv},~L. 1996, IBVS, 4368
\bibitem[2007]{mukai07} Mukai,~K., Ishida,~M., Kilbourne,~C. et al.
  2007, PASJ, 59, 177
\bibitem[1996]{munari96} Munari,~U., Yudin,~B.~F., Kolotilov,~E.~A.,
  Tomov,~T.~V. 1996, A\&A, 311, 484
\bibitem[1997]{munari97} Munari,~U., Renzini,~A., Bernacca,~B.~L. 1997,
  Hipparcos--Venice '97 (ESA SP-402), 413
\bibitem[2002]{munari02} Munari,~U., Zwitter,~T. 2002, A\&A, 383, 188
\bibitem[1994]{robinson} Robinson,~K., Bode,~M.~F., Skopal,~A., Ivison,~R.~J.,
  Meaburn,~J. 1994, MNRAS, 269, 1
\bibitem[2007]{shugarov07} Shugarov,~S.~Yu., Tatarnikova,~A.~A.,
 Kolotilov,~E.~A., Shenavrin,~V.~I., Yudin,~B.~F. 2007, BaltA, 16, 23
\bibitem[1995]{skopal95} Skopal,~A. 1995, IBVS, 4157
\bibitem[1997]{skopal97} Skopal,~A. 1997, IBVS, 4495
\bibitem[1998]{skopal98} Skopal,~A. 1998, CoSka, 28, 87
\bibitem[1996a]{skopal96a} Skopal,~A., Bode,~M.~F., Bryce,~M., et al.
  1996a, MNRAS, 282, 327
\bibitem[1996b]{skopal96b} Skopal,~A., Bode,~M.F., Lloyd,~H.~M.,
  Tamura, S.~1996b, A\&A, 308, L9
\bibitem[1995]{skopal95jt} Skopal,~A., Hric,~L., Chochol, D., et al.
  1995, CoSka, 25, 53
\bibitem[2004]{skopal04} Skopal,~A., Pribulla,~T., Va\v{n}ko,~M.,
  et al. 2004, CoSka, 34, 45
\bibitem[2000]{skopal00} Skopal,~A., Pribulla,~T., Wolf,~M.,
  et al. 2000, CoSka, 30, 29
\bibitem[2002]{skopal02} Skopal,~A., Va\v{n}ko,~M., Pribulla,~T.,
  et al. 2002, CoSka, 32, 62
\bibitem[2007]{skopal07} Skopal,~A., Va\v{n}ko,~M., Pribulla,~T.,
  et al. 2007, AN, 328, 909
\bibitem[1978]{slovak78} Slovak,~M.~H., Africano,~J. 1978, MNRAS, 185, 591
\bibitem[2003]{soko03} Sokoloski,~J.~L., Kenyon,~S.~J. 2003, ApJ,
  584, 1027
\bibitem[1987]{solf87} Solf,~J. 1987, A\&A, 180, 207
\bibitem[2006]{taranova06} Taranova,~O., Shenavrin,~V. 2006, PZP, 6, 28
\bibitem[1986]{taylor86} Taylor,~A.~R., Seaquist,~E.~R., Mattei,~J.~A.
  1986, Natur, 319, 38
\bibitem[1993]{winckel93} van Winckel,~H., Duerbeck,~H.~W., Schwarz,~H.~E.
  1993, A\&AS, 102, 401
\bibitem[2006]{wheatley06} Wheatley,~P.~J., Kallman,~T.~R. 2006, MNRAS,
  372, 1602
\bibitem[1979]{yamashita79} Yamashita,~Y., Maehara,~H. 1979, PASJ, 31, 307
\end{thebibliography}
\end{document}